\input harvmac
\Title{ \vbox{\baselineskip12pt
\hbox{hep-th/0004095}
\hbox{HUTP-00/A011}
\hbox{NUB-3209}}}
{{\vbox{
\centerline{Branes and Fluxes in $D=5$ Calabi-Yau}\bigskip
\centerline{Compactifications of M-Theory}}}}\
\smallskip
\centerline{Micha{\l} Spali\'nski}
\smallskip
\centerline{\it Jefferson Laboratory of Physics}
\centerline{\it Harvard University, Cambridge, MA 02138, USA}\bigskip
\centerline{Tomasz R. Taylor}
\smallskip
\centerline{\it Department of Physics, Northeastern University}
\centerline{\it Boston, MA 02115, USA}
\bigskip

\bigskip

\def\tilde{\widetilde}

\def\pl{Phys. Lett. B~}
\def\np{Nucl. Phys. B~}
\def\pr{Phys. Rev. D~}
\def\prl{Phys. Rev. Lett.~}
\def\tilde{\widetilde}

\def\half{{1\over 2}}

\def\sixth{{1\over 6}}
\def\prepot{{\cal V}}
 
\vskip .3in
We discuss Poincar\'e three-brane solutions in $D=5$ M-Theory 
compactifications on Calabi-Yau (CY) threefolds with $G$-fluxes.
We show that the vector moduli freeze at an attractor point.
In the case with background flux only, the spacetime geometry contains
a zero volume singularity with the three-brane
and the CY space shrinking simultaneously to a point. This problem can be
avoided 
by including explicit three-brane sources.
We consider  two cases in detail: a single brane and,
when the transverse dimension is compactified on a
circle, a pair of branes with opposite tensions. 
\def\underarrow#1{\vbox{\ialign{##\crcr$\hfil\displaystyle
 {#1}\hfil$\crcr\noalign{\kern1pt\nointerlineskip}$\longrightarrow$\crcr}}}
\Date{April 2000}

\newsec{Introduction and Summary}

One of the most challenging problems in string theory is to uncover
the vacuum selection mechanism. Assuming that this mechanism can be 
understood within the framework of low-energy field theory, the problem
amounts to the computation of the effective potential for the moduli fields.
For quite a long time, non-perturbative effects, like gaugino condensation, 
have been considered (with limited success)
as a possible source of such potentials. More recently, starting with the
work 
of Polchinski and Strominger \ref\post{J. Polchinski and A. Strominger,
\pl 388 (1996) 736, hep-th/9510227.}, the research focus has shifted 
to compactifications involving\nref\jm{J. Michelson, \np 495 
(1997) 127, hep-th/9610151.}\nref\gvw{S. Gukov, C. Vafa, E. Witten,
hep-th/9906070.}\nref\das{K. Dasgupta, G. Rajesh, S. Sethi, JHEP 9908 (1999)
023, hep-th/9908088.}\nref\guk{S. Gukov, hep-th/9911011.} 
non-vanishing background fluxes of various antisymmetric tensor 
fields -- so-called $G$-fluxes \refs{\jm-\guk}. In fact, for Calabi-Yau 
compactifications of type II theory, $G$-fluxes can generate \ref\tomek{T.R.
Taylor and C. Vafa, \pl 474 (2000) 130, hep-th/9912152.}\ 
the most general form  of the (super)potential allowed by $N=2$ supersymmetry.
In five dimensions, in Calabi-Yau compactifications
of M-Theory, $G$-fluxes of the eleven-dimensional three-form gauge field 
produce a similar potential \nref\lukas{A. Lukas, B.A. Ovrut, K.S. Stelle, 
D. Waldram, \pr 59 (1999) 086001, hep-th/9803235.}
\nref\burt{A. Lukas, B.A. Ovrut, K.S. Stelle, 
D. Waldram, 
\np 552 (1999) 246, hep-th/9806051.}\nref\bg{K. Behrndt and S. Gukov, 
hep-th/0001082.}\refs{\lukas,\burt,\bg}. 
This type of compactification is particularly interesting, 
since M-Theory provides a powerful setup for studying string dynamics.

It is interesting to look at so-called warped compactifications with
Poincar\'e invariance in this context. Since string theory is
known to contain higher dimensional extended objects in an essential way, it
is natural to look at compactifications which involve them in a nontrivial
fashion. Recently, some examples of this type have been studied
in connection with the hierarchy problem 
\ref\rs{L. Randall and R. Sundrum, \prl 83 (1999) 3370, hep-th/9905221;
\prl 83 (1999) 4690, hep-th/9906064.}\
and with the cosmological constant problem \ref\cosmo{N. Arkani-Hamed,
S. Dimopoulos, N. Kaloper, R. Sundrum, hep-th/0001197; S. Kachru, M. Schulz,
E. Silverstein, hep-th/0001206, hep-th/0002121; 
S.P. de Alwis, hep-th/0002174; 
S. Forste, Z. Lalak, S. Lavignac, H.P. Nilles,
hep-th/0002164.}. 

In this work, we study the classical field equations of M-Theory compactified
from  $D=11$ to $D=5$ on Calabi-Yau (CY) threefolds with various 
$G$-flux configurations. In the absence of fluxes, the effective field
theory is $D=5$ supergravity \ref\gun{M. G\"unaydin, G. Sierra, 
P.K. Townsend, \np 242 (1984) 244;
\np 253 (1985) 573.}\ coupled to a number of vector and hyper
multiplets
\nref\cadav{A.C. Cadavid, A. Ceresole, R. D'Auria, S. Ferrara,
\pl 357 (1995) 76, hep-th/9506144.} 
(as determined by the cohomology of the Calabi-Yau space \cadav).
The presence of fluxes results in gauged supergravity \gun\ with
a non-vanishing potential \refs{\lukas,\burt,\bg}.

We first discuss the case of  smooth background fluxes, i.e.\ 
without explicit sources. We consider a Poincar\'e three-brane solution of
the form 
\eqn\metric{
ds^2=e^{2\phi(u)}\eta_{\alpha\beta}dx^{\alpha}dx^{\beta} + du^2\ ,
}
where the $x$-coordinates parameterize
the $D=4$ three-brane world-volume, $\eta_{\alpha\beta}$ is the flat 
four-dimensional
Minkowski metric, while $u$ is the ``fifth'' coordinate (transverse to the
three-brane).
The Weyl factor $e^{2\phi(u)}$ depends only on the transverse coordinate
$u$ and is related by the field equations to the CY volume.
The solution \metric\ exhibits a zero volume singularity with
the three-brane and the CY threefold shrinking simultaneously to a point.
On the other hand, the shape of the Calabi-Yau manifold, determined by its
vector (K\"ahler) moduli, remains frozen at a point
corresponding to the extremal value of the central charge. In fact,
the stability condition turns out to be exactly the same as the attractor
equation \nref\fk{S. Ferrara and R. Kallosh, \pr 54 (1996) 1514,
hep-th/9602136; \pr 54 (1996) 1525, hep-th/9603090.}\nref\attr{A. Chou,
R. Kallosh, J. Rahmfeld, S.J. Rey, M. Shmakova, W.K. Wong, \np 508 (1997) 147,
hep-th/9704142.}\refs{\fk,\attr}\
for a $D=5$ black hole, with the charges identified as $G$-fluxes.

The zero volume singularity can be avoided by introducing 
a $G$-flux discontinuity across a 
three-brane source. 
While the detailed interpretation of the sources is beyond the scope of 
this paper, the required tension (determined by the equations of motion)
indicates the presence of a fivebrane wrapped on a Calabi-Yau two-cycle. 
In the case of a compact transverse dimension, 
we construct flux configurations supported entirely by a pair of effective
brane sources with opposite tensions.
This system is somewhat similar to the one considered by Randall and 
Sundrum \rs.
In the present case, however, the bulk spacetime is not AdS. 

The paper is organized as follows. In Section 2, we establish notation
and review $D=5$ M-Theory CY compactifications with $G$-fluxes.
The Poincar\'e three-brane solution is presented in Section 3. 
In Section 4, we establish the connection with the attractor mechanism.
We introduce flux sources in Section 5. In Section 6, we examine
the supersymmetry variations of fermions and identify the unbroken 
supersymmetry
transformations. Section 7 contains conclusions and outlook.

\newsec{CY Compactification of M-Theory with Background Fluxes}
\nref\aft{I. Antoniadis, S. Ferrara, T. R. Taylor,
\np 460 (1996) 489, hep-th/9511108.}

This section is a brief review aimed at fixing notation. The
compactification of $D=11$ supergravity on a Calabi-Yau 
threefold with Hodge numbers $(h_{1,1},h_{2,1})$ results in an $N=2$, $D=5$
supergravity theory interacting with $h_{1,1}-1$ vector multiplets and
$h_{2,1}+1$ hypermultiplets \cadav. In our discussion, hypermultiplets
play no role, except for the universal hypermultiplet involving the CY
volume. 
The relevant part of the action is determined by a 
cubic prepotential $\prepot$ which is fixed by the CY intersection
numbers. Details of this can be found in a number of  
references, see e.g.\ \refs{\cadav,\aft}. 
Modifications arising from the presence of 
background fluxes of the four-form field-strength have been discussed in
\refs{\lukas,\burt}\ and more recently 
recently in \bg. 
The presence of a background flux implies
that the  supergravity is gauged \gun, and a potential of a specific 
form is induced in the five-dimensional effective action. 

Let us denote by $M^i$ the scalar  K\"ahler moduli of the Calabi-Yau threefold 
so that the K\"ahler form
\eqn\kal{
J=M^i\omega_i\ ,
}
where $\omega_i,~i=1,\dots,h_{1,1}$ is a basis of $H^{(1,1)}$ two-forms,
and the CY volume
\eqn\prep{
\prepot(M)={1\over 6}\int_{CY}J\wedge J\wedge J=
{1\over 6} c_{ijk} M^i M^j M^k\ , 
}
where $c_{ijk}$ are the intersection numbers.
In the absence of background fluxes the action\foot{In our 
conventions, the metric has signature $(-++++)$ and the
Ricci tensor $R_{\mu\nu}=\partial_{\rho}\Gamma^{\rho}_{\mu\nu}+
\dots$}\ is given by \cadav\ 
\eqn\cad{
S_0 = \int d^5 x \sqrt{-g} \big[\half\prepot R + \half (\prepot G_{ij} +  
\partial_i\partial_j\prepot)\ \partial_{\mu} M^i \partial^{\mu} M^j
+\dots\big]}
where the moduli space metric is
\eqn\modmetr{
G_{ij}(M) = {i\over 2\prepot}\int_{CY}\omega_i\wedge {^{\star}}\omega_j=
-\half\partial_i\partial_j\ln\prepot\ .}
In the above equations, $\partial_i\equiv{\partial\over\partial M^i}$.
It is often
convenient to parameterize the moduli space in a way that makes
manifest the decoupling of vector multiplets and hypermultiplets. This
entails a Weyl rescaling of the metric as well as 
introducing the special coordinates  $X^i= M^i\prepot^{-1/3}$,
and treating the volume \prep\ as an independent field belonging to the
universal hypermultiplet. For the present purpose this is not so
useful, so the volume $\prepot$ will be regarded as a function of the moduli
as given in \prep. 

The presence of background fluxes gives rise to a potential
\refs{\lukas,\burt,\bg}.
We will consider the following four-form field strength of the 
three-form gauge field:
\eqn\flux{
G_{\rm flux}={i\over 2\prepot}\alpha_iG^{ij}\ {^{\star}}\omega_j\ , 
}
where $\alpha_i$ are integers, as required by flux quantization conditions,
and $G^{ij}$ is the inverse of the moduli space metric \modmetr. 
The scalar potential originates from $D=11$ kinetic terms, which upon
compactification yield the term
\eqn\pot{
S_{\rm flux} = - {1\over 8}\int d^5 x \sqrt{-g}\ \prepot^{-1}
 G^{ij}\alpha_i\alpha_j\ . 
}

Equations \cad\ and \pot\  are written in the string frame. 
In order to obtain the canonical Einstein-Hilbert term one performs
the Weyl rescaling
\eqn\xxx{ds^2_{\rm E}=\prepot^{2/3}ds^2\ .}
In the Einstein frame, the full action, $S=S_0+S_{\rm flux}$, reads
\eqn\action{
S = \int d^5 x \sqrt{-g} \big[\half R - \half G_{ij}\partial_{\mu} M^i
\partial^{\mu} M^j -\sixth \partial_{\mu} (\ln \prepot) \partial^{\mu}
 (\ln \prepot) - 
{1\over 8}\prepot^{-8/3} G^{ij}\alpha_i\alpha_j +\dots\big] \ .
}

\newsec{The Solution}

In this section, we solve the classical field equations for the
extremum of the action \action.   We look for a gravitational
background of the form \metric\ representing a Poincar\'e-symmetric
three-brane in  five dimensions.
The non-vanishing components of the corresponding Einstein tensor are
\eqn\einst{\eqalign{
E_{\alpha\beta}&= \eta_{\alpha\beta} e^{2\phi} [3\phi^{\prime\prime} + 6
(\phi^\prime)^2 ]\ , \cr
E_{uu} &= 6 (\phi^\prime)^2\ ,}}
where the prime denotes a derivative with respect to the transverse
coordinate $u$. 

The initial observation is that
the variation of the action with respect to the moduli $M$ contains the terms
\eqn\varact{
\delta S = \int d^5 x \sqrt{-g} [ - \half \partial_{\mu} M^i
\partial^{\mu} M^j\delta G_{ij} + {1\over 8}  
\prepot^{-8/3}\alpha_i\alpha_j  G^{im}
G^{jn}\delta G_{mn} + \dots] 
}
which suggests considering 
solutions with moduli depending only on $u$
(in line with the Poincar\'e symmetry of the metric \metric), together 
with a BPS-like Ansatz 
\eqn\ansatz{
2(M^i)' =\prepot^{-4/3} G^{ij}\alpha_j\ .
}
This Ansatz leads to several simplifications. First of all,
the $(uu)$ component of Einstein's equations becomes
\eqn\eineq{
6(\phi')^2 = \sixth ({d\ln\prepot\over du})^2\ .
}
This is solved by
\eqn\weyl{
e^{2\phi}=e^{2\phi_0}\ \prepot^{1/3}\ ,
}
where $\phi_0$ is a constant. 
We will ignore the second solution, $e^{2\phi}\propto\prepot^{-1/3}$, 
since it fails to satisfy some other field 
equations; we will comment on this below. 
The $(\alpha\beta)$ components of Einstein's equations simplify after
using Eqs. \ansatz\ and \weyl. They become: 
\eqn\eqtwo{
(\prepot^{2/3})'' +{2\over3} \prepot^{-2/3}(\alpha_iM^i)'=0\ .
}

The remaining terms in the variation of the action with respect to the
moduli, after substituting Eqs.\ansatz\ and \weyl, lead to the
following equations:
\eqn\mod{
(\prepot^{-2/3})'\alpha_i+\prepot^{-1}(\prepot^{2/3})''
{\partial_i\prepot}+{4\over 3}\prepot^{-5/3}(\alpha_kM^k)'
{\partial_i\prepot}=0\ .
}
For this to have a solution it must be the case that $\alpha_i$ is parallel
to $\partial_i\prepot$.  
Thus it is natural to look for a solution such that
\eqn\furtans{
\alpha_i =  {\zeta\over 3}\prepot^p\partial_i\prepot\ ,
}
where $\zeta$ and $p$ are constants. It would seem that the two Ansatze
\ansatz\ and 
\furtans\ impose too many constraints; fortunately, this is not the case.
First, by checking the compatibility of Eq.\furtans\ with Eqs.\ansatz\ and 
\eqtwo\
we find\foot{This is the point where the second
solution of Eq.\eineq\ fails to be compatible.} that the power $p=-2/3$.  
In this way, Eq.\furtans\
becomes
\eqn\furt{\prepot^{-2/3}\partial_i\prepot={3\alpha_i\over\zeta } \ .}
In the process, we also solve the second Einstein equation \eqtwo,
with the result:
\eqn\volsol{\prepot=\prepot_0+\zeta u\ ,}
where $\prepot_0$ is a constant. Finally, we use Eqs.\furt\ and \volsol\
to verify that the moduli equation of motion \mod\ is indeed satisfied.
In this way, Eqs.\furt\ and \volsol\ together with Eq.\weyl\ 
yield a consistent solution of all field equations. 

A few remarks are in order here. Note that by using the formulae of
the previous section, Eq.\furt\ can be rewritten in
a more geometric way as an equation\foot{We are grateful to C. Vafa for
pointing this out.} describing the flux \flux:
\eqn\fluxx{
G_{\rm flux}={\zeta\over 6}\prepot^{-2/3}J\wedge J\ .
}
Furthermore, since Eq.\furt\ is invariant under the rescaling
$M^i\to\lambda M^i$ with an arbitrary constant $\lambda$, it is possible
to rewrite it exclusively in terms of the special coordinates
\eqn\exi{
X^i(M)=M^i\prepot^{-1/3}\ ,
}
which satisfy the constraint
\eqn\nux{
\prepot(X)={1\over 6} c_{ijk} X^i X^j X^k=1\ . 
}
Recall that special coordinates parameterize the $(h_{1,1}{-}1)$-dimensional
vector multiplet space \gun.
In this way, one obtains
\eqn\cxx{c_{ijk}X^j X^k={6\alpha_i\over\zeta } \ .}
The above equations freeze $h_{1,1}$ special coordinates
at constant vacuum expectation values\foot{Of course, this is provided that
a solution exists.}\ depending on the intersection 
numbers, fluxes and the constant $\zeta$. The latter  is not independent:
$\zeta$ can be expressed in terms of the intersection numbers and fluxes 
by using the constraint \nux. For the purpose of illustration, we discuss
below two simple examples.

{\it Example 1:} $h_{1,1}=1, ~\prepot(S)=S^3$. This is a model without
vector multiplets, for example a quintic CY. There is a trivial solution
\eqn\exone{
X^S=1\qquad,\qquad \zeta=\alpha_S\ .
}

{\it Example 2:} $h_{1,1}=2, ~\prepot(S,T)=ST^2-{1\over 3} T^3$.
$X_{12}(1,1,2,2,6)$ CY with \nref\witt{E. Witten, \np 471 (1996) 195,
hep-th/9603150.}one vector multiplet and a flop transition \refs{\aft,\witt}. 
A simple calculation yields
\eqn\extwo{
X^S ={\zeta (\alpha_S+\alpha_T)\over 3\alpha_S
(\alpha_T-\alpha_S)}\ ,\qquad
X^T = {2\zeta \over 3(\alpha_T-\alpha_S)}\ ,\qquad
\zeta={3\over 2^{2/3}}\alpha_S^{1/3}(\alpha_T-\alpha_S)^{2/3}\ .}
For generic fluxes $\alpha_T>\alpha_S>0$, 
this is a regular solution valid in the K\"ahler cone $S>T$. 
However, if $\alpha_T=\alpha_S$, it is pushed to the flop at $S=T$. 

The above solution has previously been obtained in \burt\ (although 
written in a different parameterization) by solving the supersymmetric
Killing equations with constant vector moduli,
$(X^i)^\prime=0$.\foot{Ref.\burt\ 
contains also a class of solutions involving $u$-dependent moduli.} Our
derivation utilizes the field equations and yields the same result,
although the starting point, Eq.\ansatz, is a weaker Ansatz than
$(X^i)'=0$.  We will be using these field equations in Section 5 to obtain
some information on the tension of three-brane sources, without assuming
that these sources preserve bulk supersymmetry.

It is also worth mentioning that Eq.\cxx\ has a nice interpretation in terms 
of very special
geometry: the surface $\prepot (X)=1$ tends to align in such a way that 
its normal 
vector becomes parallel to the flux vector $\alpha_i$ \bg.

To summarize, we obtain a Poincar\'e 
three-brane solution which,
for generic values of background fluxes, freezes the vector moduli fields
at constant vacuum expectation values, fixing the shape of Calabi-Yau
manifold.  On the other hand, the hypermultiplet modulus that determines
the volume becomes a linear function of the transverse
coordinate $u$, see Eq.\volsol. There is an inevitable singularity at 
$u=-\prepot_0/\zeta$, where  the Calabi-Yau manifold
shrinks to a point.
Then the three-brane Weyl factor also vanishes, see Eq.\weyl,
therefore the whole $D=10$ spacetime collapses to one point.
We will be revisiting this problem later.

\newsec{The Attractor Connection}
\nref\sabra{W. A. Sabra, Mod. Phys. Lett.
A 13 (1998) 239, hep-th/9708103.}

It is well known \refs{\fk,\attr,\sabra} that the entropy of
five-dimensional BPS black hole solutions of $N=2$ Einstein-Maxwell
supergravity  
is determined by the extremal value of the central charge.
This value is attained at the horizon which from
the point of view of the vector moduli space acts as an attractor point. 
Such black holes appear in CY compactifications
of M-Theory and 
their entropy can be computed at the microscopic level
by counting the number of M2-branes wrapping around CY two-cycles 
\ref\vafa{C. Vafa, Adv. Theor. Math. Phys. 2 (1998) 207, hep-th/9711067.}.
We will now show that in the Poincar\'e three-brane solution 
the vector moduli are frozen 
at exactly the same attractor point, with the fluxes $\alpha_i$
identified as BPS charges.

First, note that homogeneity of the volume, Eq.\prep, together with 
Eq.\furt\ imply  that
\eqn\zet{
\zeta=\alpha_iX^i(M)= Z[X(M)]\ ,
}
where $Z$ is the central charge for a BPS state with electric
charges $\alpha_i$ \aft. Given this, Eq.\furt\ 
can be written as 
\eqn\ext{
{\partial Z \over\partial M^i}[X(M)]=0\ .
}
This means that the vector moduli $X^i$ are frozen at the extremum 
of the central 
charge. It is also clear that the constant $\zeta$ 
is equal to the extremal value of the central charge. The above equations
can be rewritten in terms of very 
special geometry, without referring to
the underlying moduli $M$, as the familiar \attr\ $D=5$ attractor stability 
condition:
\eqn\diz{D_i Z=0\qquad, \qquad Z(X)=\alpha_iX^i \qquad,
\qquad \zeta=Z\big|_{D_i Z=0}\ ,}
where we used the covariant derivative
\eqn\ddef{D_i={\partial\over \partial X^i}-{1\over 6}c_{ijk}X^jX^k}
appropriate for $Z(X)$ defined on the surface $\prepot (X)=1$.

We conclude that in the presence of a Poincar\'e three-brane,
the vector moduli are forced to the same attractor configuration
as on the horizon of a charged black hole. This indicates that
charged black holes may play an important role in resolving the zero volume
singularity.

\newsec{Three-brane Sources}

We can avoid the singularity if we consider\nref\bov{A. Lukas, B.A. Ovrut,
D. Waldram, \pr 59 (1999) 106005, hep-th/9808101.}
other flux configurations. We will discuss
fluxes jumping across one or two 
three-brane sources. Their tension will be determined below. 
Similar ideas have
been discussed before in various contexts in several places, including
\refs{\bov,\gvw,\tomek,\bg}.
We consider the cases of non-compact and compact transverse dimension
separately.

\subsec{Non-compact transverse dimension: a single brane source}

Let us consider a three-brane located at $u=0$, with the flux jumping
from $-\alpha_i$ for $u<0$ to $+\alpha_i$ for $u>0$. The scalar potential 
does not change upon reversing the flux direction, therefore the bulk action
remains the same as in Eq.\action. Similarly, the solution of Section 3
remains valid for $u>0$. 
In order to obtain
a solution for $u<0$ it is sufficient to change the signs
$\alpha_i\to-\alpha_i$ and $\zeta\to-\zeta$. 
Hence the moduli remain frozen at the same attractor point as before,
see \cxx, and the Weyl factor is still given by Eq.\weyl.
On the other hand, the CY volume\foot{The additive
integration constants have been adjusted to ensure that the volume
is a continuous function of $u$.}
\eqn\zig{
\prepot=\prepot_0+\zeta |u|\ .
}
It is clear that the zero volume singularity can indeed be avoided if $\zeta>0$
(for $\zeta<0$ one would have to reverse the flux directions).

The cusp at $u=0$ contributes additional terms proportional to $\delta (u)$
to the field equations. Therefore, in order
to obtain a self-consistent solution valid everywhere in $D=5$ spacetime,
the bulk action must be supplemented by a term of the form
\eqn\sbrane{
S_{\rm brane}= - \int d^4 x \sqrt{-g^{(4)}}f(M)\ ,
}
representing an explicit three-brane source with efective tension $f(M)$ at
$u=0$. 
The  four-dimensional metric
$g^{(4)}_{\alpha\beta}$ is induced by the
bulk metric $g_{\mu\nu}$: 
$g^{(4)}_{\alpha\beta}\equiv
\delta_{\alpha}^{\mu}\delta_{\beta}^{\nu}g_{\mu\nu}(u=0)$.
The moduli-dependent three-brane tension $f(M)$ is constrained by the field
equations in the following way. The $(\alpha\beta)$ components of Einstein's
equations require 
\eqn\fein{
f= - {\zeta\over\prepot_0}\ .
}
On the other hand, the moduli field equations dictate
\eqn\fzwei{
{\partial f\over\partial M^i}= 3\prepot_0^{-4/3}\alpha_i\ .
}
It is easy to find a function that, with the help of Eqs.\furtans\ and \zet,
satisfies these constraints:
\eqn\fsol{
f(M)= - \prepot^{-4/3}\alpha_iM^i\ .
}
If one returns to the string frame
by undoing the Weyl rescaling \xxx\ so that $\sqrt{-g^{(4)}}\to
\prepot^{4/3}\sqrt{-g^{(4)}}$, then the three-brane tension becomes
\eqn\fstring{
f_s(M)= - \alpha_iM^i\ = - \int_{[G]}J\  ,
}
where $[G]$ is the two-cycle Poincar\'e dual to the four-form
field strength $G_{\rm flux}$ of Eq.\flux:
\eqn\gcycle{
\int_{[G]}\omega_i=\int_{CY}G_{\rm flux}\wedge\omega_i
=\alpha_i\ .
}
Note that the required tension is negative. Its magnitude however is equal
to the tension of a fivebrane wrapping a CY two-cycle. The string
theory origin of 
such objects is likely to be found in the framework of F-theory
compactifications \ref\verlin{C.S. Chan, P.L. Paul, H.
Verlinde, hep-th/0003236}.

\subsec{Compact transverse dimension: a pair of branes with opposite
tension} 

Let us assume that the transverse dimension is compactified on a circle,
with $u\in (-1,1]$. Starting from the non-compact domain wall solution 
discussed before,
we can construct a simple periodic configuration with the flux changing
direction ($\alpha_i\to-\alpha_i$) at $u=0$ and then reversing back to its 
original value at $u=1$. The CY volume $\prepot(u)$, Eq.\zig,
now becomes a periodic function zigzagging between $\prepot_0$ and 
$\prepot_0+\zeta$. The additional cusp at $u=1$ forces
us to introduce another flux source. By repeating the previous arguments
one can identify this source as a brane with the tension
$\tilde{f}=-f$. This brane could be identified with an M-theory fivebrane
wrapping a CY two-cycle. 
In this way, the pair of branes with opposite tension supports
a flux configuration in the compactified space.

This solution is somewhat similar
to the configuration studied by Randall and Sundrum \rs.
There is however no room in M-Theory for a
fine-tuning of cosmological constants: 
the bulk vacuum energy originates from $G$-fluxes while tensions of the
effective three-brane sources are determined by the Calabi-Yau geometry. 
As a result, one obtains a Weyl factor which is different
from AdS-like exponential warp factors that localize gravity.
Furthermore, one would expect that the equilibrium of
brane configurations considered here is not stable under small 
perturbations\foot{In principle, this problem could be circumvented
by working on an
orbifold $S^1/Z_2$ and placing the sources at the fixed
points.}. All these points deserve further investigation.

\newsec{Supersymmetry}

In this Section, we examine the supersymmetry transformations in order
to determine what (if any) type of supersymmetry is preserved by our 
solutions. To that end, it is convenient to use
the notation of \ref\bagger{R.
Altendorfer, J. Bagger, D. Nemeschansky, hep-th/0003117.}, 
with the two $N=2$ supersymmetry generators
labelled by $\pm$. 
In the gravitational background \metric, the non-vanishing
components of the spin connection can be rewritten using Eq.\weyl\ as
\eqn\conn{
\omega^{au}=(\prepot^{1/6})'dx^a\ ,
}
where $a$ denotes the $D=4$ Lorentz indices.
Thus the supersymmetry variations of the gravitinos become
\eqn\gra{\eqalign{
\delta\psi^{\pm}_{\alpha}&=2\partial_{\alpha}\eta^{\pm}
\pm i{\zeta\over 6\prepot} \sigma_{\alpha}\bigg( {\prepot^\prime 
\over\zeta}\bar{\eta}_{\mp}+\bar{\eta}_{\pm}\bigg) \ , \cr
\delta\psi^{\pm}_{u}&=2\partial_{u}\eta^{\pm}+{\zeta\over
6\prepot}\eta^{\mp} \ ,} 
}
where we used Eq.\zet. 
In order to find the unbroken supersymmetries we first set these variations 
to zero and solve the corresponding Killing spinor equations.

For the solution of Section 3, i.e.\ in the absence of brane sources,
$\prepot^\prime =\zeta$, and the Killing equations are solved by
\eqn\kil{
\eta_0^{+}= - \eta_0^{-}=\epsilon\prepot^{1/12}\ ,
}
where $\epsilon$ is a constant Weyl spinor. 

If a source is inserted
at $u=0$, as in the examples discussed in Section 4, then 
$\prepot^\prime ={\rm sgn}(u)\zeta$, and
\eqn\kilm{
\eta^{+}=\epsilon\prepot^{1/12}\qquad ,\qquad
\eta^{-}= - \epsilon\prepot^{1/12}{\rm sgn}(u) \ .
}
However, in this case 
\eqn\psu{
\delta\psi^{-}_{u}=4\epsilon\prepot_0^{1/12}\delta(u)\ ,
}
hence the Killing equations are satisfied everywhere in the bulk
but they are {\it not\/} satisfied on the brane hypersurface.
Furthermore, it is easy to see that for the respective solutions,
the spinors \kil\ and \kilm\  give vanishing
supersymmetry variations of all other fermions: hyperinos and gauginos.
In particular, the gaugino variations vanish for the moduli
frozen at the attractor point \diz.

In this way, we reach the conclusion that 
the singular solution preserves $N=1$ supersymmetry.
The regular solutions involving brane sources preserve $N=1$ supersymmetry 
in the bulk, however they break it on the branes as long as $\prepot_0\neq
0$. 

\newsec{Conclusions and Outlook}

In this work we studied $D=5$ Calabi-Yau compactifications
of M-Theory with background $G$-fluxes and explicit effective three-brane  
sources. 
In the absence of sources there exists an $N=1$ supersymmetric solution
with the metric representing a Poincar\'e three-brane. The Weyl factor
depends on the transverse coordinate $u$ as $(\prepot_0+\zeta u)^{1/3}$.
At $u=-\prepot_0/\zeta$ the three-brane as well as the CY manifold 
shrink to a point. The vector moduli that
determine the shape of the Calabi-Yau manifold
remain frozen at a point similar to the well-known black hole attractor
point, which suggests that charged black holes
play an important role in resolving the zero volume singularity.

The singularity can be avoided altogether by introducing a  
$G$-flux source along the three-brane hypersurface. In this case, the CY 
volume 
reaches its minimum at the position of the source. 
Although the string theory origin of such a source remains unclear, its  
tension is equal (up to a sign) to the volume of the two-cycle dual to the 
flux vector $\alpha_i$. 
If the transverse dimension is
compactified on a circle, one can also construct a brane
configuration similar to the Randall-Sundrum configuration. In M-Theory
though, 
the bulk vacuum energy is completely 
determined by $G$-fluxes and the brane tensions
by CY geometry -- as a result one obtains a gravitational background which
is 
not of the AdS-type.

There is a number of points deserving further investigation.  We raised
several stability issues. In the solutions involving three-brane sources,
the minimum value of CY volume remains undetermined at the level of
classical field equations. The obvious question is how this zero mode, and
the solution in general, are affected by quantum corrections. Furthermore,
one should analyze the stability of brane 
configurations with respect to their relative positions. It is possible
that answers to these questions can be obtained in the framework of
F-theory (or its orientifold limits).

\bigskip\noindent
{\bf Acknowledgements}~ We have benefited from discussions
with Costas Bachas, Michael Gutperle, Ashoke Sen and Cumrun Vafa. The
work of T.R.T. was supported in  
part by NSF grant PHY-99-01057 and that of
M.S. was supported in part by NSF grant PHY-98-02709.

\listrefs
\end